\documentclass[twocolumn,nofootinbib,preprintnumbers]{revtex4-1}
\usepackage{graphicx}
\usepackage{amsmath, amsthm, amssymb}
\usepackage{slashed}
\usepackage{url}
\usepackage[pdftex]{hyperref}
\usepackage{color}
\usepackage{bm}
\usepackage{mathtools}

\newcommand{\bea}{\begin{eqnarray}}
\newcommand{\eea}{\end{eqnarray}}

\begin{document}

\preprint{UCI-HEP-TR-2019-11}

\title{An Emergent Solution to the Strong CP Problem}
\author{Jason Arakawa}
\author{Arvind Rajaraman}
\author{Tim M.P. Tait}
\affiliation{Department of Physics and Astronomy, University of
California, Irvine, CA 92697-4575 USA}
\date{\today}
\begin{abstract}
We construct a theory in which the solution to the strong CP problem is an emergent property of the background
of the dark matter in the Universe.  The role of the axion degree of freedom is played by
multi-body collective excitations similar to spin-waves in the medium of the dark matter of the Galactic halo.  
The dark matter is a vector particle whose low energy interactions with the Standard Model take the form
of its spin density coupled to $G \widetilde{G}$, which induces a potential on the average spin density
inducing it to compensate $\overline{\theta}$, effectively removing CP violation in the strong sector in regions of the
Universe with sufficient dark matter density.  We discuss the viable parameter space, finding that light dark matter masses
within a few orders of magnitude of the fuzzy limit are preferred, and discuss the associated signals with this type of
solution to the strong CP problem.
\end{abstract}

\maketitle

%%%%%%%%%%%%%%%%%%%%%%%%%%%%%%%%%%%
\section{Introduction}
\label{sec:intro}
%%%%%%%%%%%%%%%%%%%%%%%%%%%%%%%%%%%

The theory of the strong interactions is well established as Quantum Chromodynamics, 
based on an SU(3)$_c$ gauge symmetry with vector-like quarks in the fundamental representation.
A wealth of observational data ranging from high energies where the theory is described as weakly coupled quarks and gluons down to low energies where
they are confined into color-neutral hadrons has established QCD as an integral building block of the Standard Model (SM).

Despite this unquestionable success, the structure of QCD contains a deep mystery: the symmetries of the theory admit a dimension four interaction term for the gluons
which violates CP:
\bea
\frac{\alpha_s}{8 \pi}~\overline{\theta} ~G_a^{\mu \nu} \widetilde{G}^a_{\mu \nu}
\eea
where $\overline{\theta} \equiv \theta + {\rm Arg}~ {\rm Det} ~M_q$ 
is the basis-independent quantity characterizing the physical combination of the strong phase $\theta$ and a phase in the quark
Yukawa interactions.  Null searches for an electric dipole moment of the neutron \cite{Afach:2015sja} require $\overline{\theta} \lesssim 10^{-10}$, in contrast to the
naive expectation that it be order 1.  While it is possible that such a tiny value is simply one of the parameters that Nature has handed us,
the extraordinarily minute experimental limit is suggestive that we explore physical explanations.

The most popular explanation invokes a fundamental axion field \cite{Wilczek:1977pj,Weinberg:1977ma,Kim:1979if,Shifman:1979if,Dine:1981rt,Zhitnitsky:1980tq},
arising as the pseudo-Nambu Goldstone boson of a spontaneous broken U(1)$_{PQ}$ symmetry \cite{Peccei:1977ur,Peccei:1977hh}
resulting in a coupling of the form
\bea
\frac{a(x)}{f_a} ~G_a^{\mu \nu} \widetilde{G}^a_{\mu \nu}.
\eea
At low scales, non-perturbative QCD dynamics induce a potential which is schematically of the form $- \Lambda^4 \cos \left( a / f_a - \overline{\theta} \right)$, inducing a vacuum
expectation value for $a$ which effectively cancels the net coefficient of the CP-violating term.  There is a vibrant experimental program underway to search for axions
in various ranges of mass \cite{Battaglieri:2017aum}.

In this article, we propose a new class of solution to the strong CP problem.
We consider a theory in which there is no fundamental axion field, but in which the dark matter, necessary
to explain cosmological observations, is composed of light vector particles which couple to the gluons in such a way that the net local spin density 
acts in some ways like an emergent degree of freedom which cancels $\overline{\theta}$.  The axion can be understood as an emergent phenomenon, similar in character to the spin-wave excitations observed in condensed matter systems.

%%%%%%%%%%%%%%%%%%%%%%%%%%%%%%%%%%%
\section{Dark Matter}
\label{sec:model}
%%%%%%%%%%%%%%%%%%%%%%%%%%%%%%%%%%%

\begin{figure}[t]
\begin{center}
\includegraphics[width=0.45\linewidth]{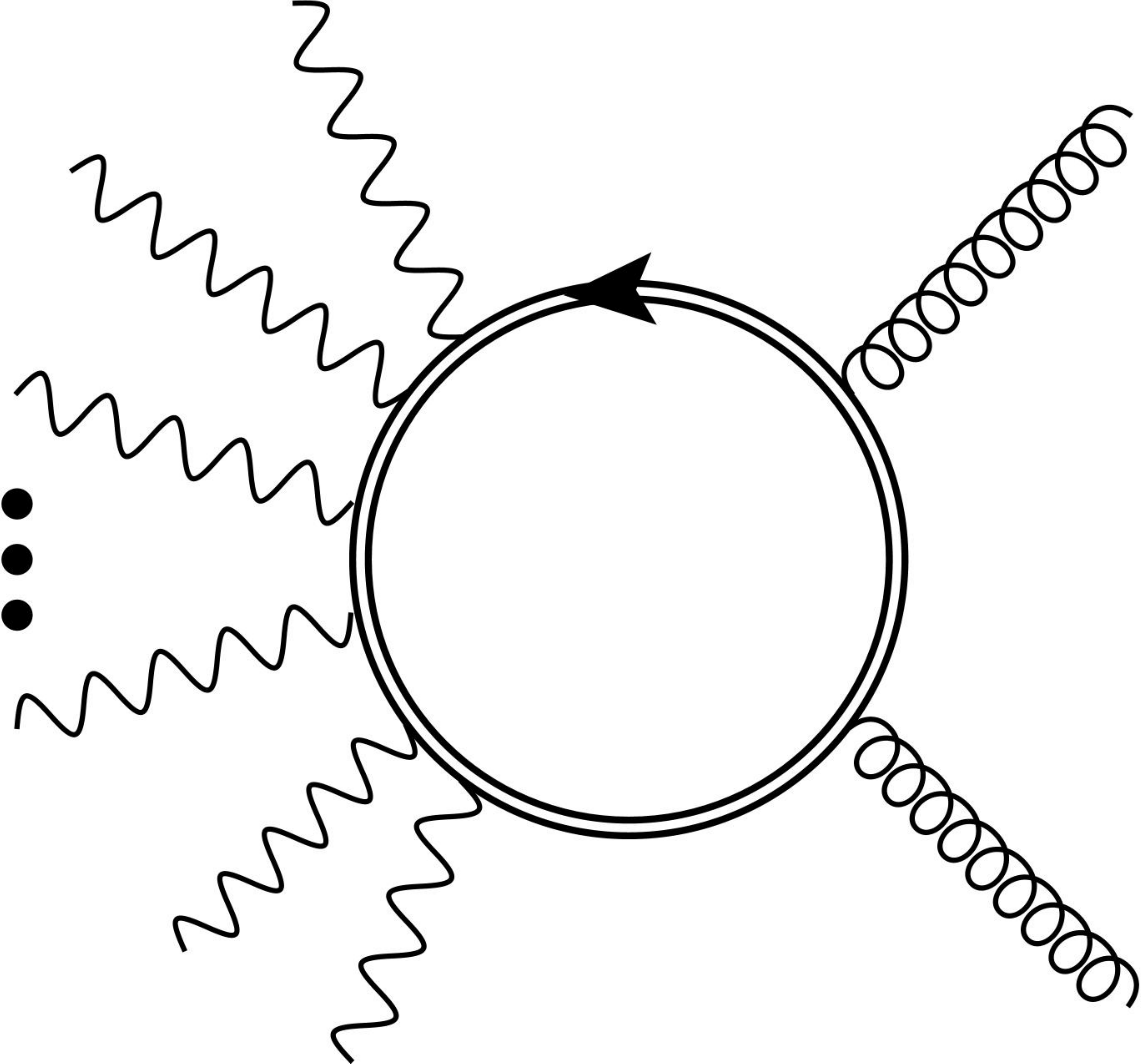}
\end{center}
\caption{Representative Feynman diagram indicating how integrating out SU(3)$_c$-charged 
fermions can generate an interaction between the dark matter
and gluons.}
\label{fig:feynman}
\end{figure}

The dark matter is assumed to be a massive vector $A_\mu$ described by the free Lagrangian,
\bea
\mathcal{L} = -\frac{1}{4}F_{\mu \nu}F^{\mu \nu} + \frac{1}{2}m^2 A_\mu A^\mu
\eea
where $F_{\mu\nu} \equiv \partial_\mu A_\nu - \partial_\nu A_\mu$ is the usual field strength tensor,
and $m$ can be understood as either a St\"uckelberg mass or as arising from a dark Higgs sector.  
We introduce an interaction between the dark matter and the SM gluons through operators of the form,
\bea
\frac{\alpha_s}{16 \pi} \frac{1}{M_*^{(6+2n)}} ~ S^{\mu \nu \rho} S_{\mu \nu \rho} \left(A^\alpha A_\alpha \right)^n ~ 
G^a_{\sigma \lambda} \widetilde{G}_a^{\sigma \lambda}
\label{eq:interaction}
\eea
where
\begin{align}
    S^{\mu \nu \rho}[A] &\equiv  F^{\mu \nu}A^\rho - F^{\mu \rho}A^\nu
    \label{eq:S}
\end{align}
is the functional of $A_\mu$ representing
the position-independent portion of the Noether current corresponding to rotations, and thus
corresponds in the non-relativistic limit to the net spin density carried by the $A_\mu$ field, $\vec{S}_i \sim \epsilon_{ijk} S^{0jk}$.
$M_*$ characterizes the strength of the interaction and has units of energy, and $n$ is an integer.  
Such interactions could be generated, for example, by integrating out heavy SU(3)$_c$-charged degrees 
of freedom which couple to the dark matter (see Figure~\ref{fig:feynman}).  
In that case, one would expect the low energy theory to contain
the whole family of operators for all values of $n$.

The interaction, Eq.~(\ref{eq:interaction}) is not manifestly gauge invariant, and can be understood to be written in the unitary gauge.
As dictated by dark gauge-invariance, the dependence of $M_*$
on the underlying UV parameters depends
on the form of the UV theory.  For example, if the SU(3)$_c$-charged 
fermions in the loop are chiral, and get their mass from the same dark Higgs vacuum expectation value 
$v_D$ which
breaks the dark gauge symmetry, one would expect 
the coefficient of the interaction to get a contribution at 1-loop of the form 
$\int d^4k~k^4 ~m_\Psi^{(2+2n)}/(k^2-m^2)^{(6+2n)}\sim \partial^4 / M_*^{6+2n}$, 
%\sim y^{2+2n} / (y v_D)^{6+2n}$, 
where $m_\Psi=y v_D$ is the mass generated by the Yukawa
interaction, and the vector fields $A_\mu$ would be longitudinal modes arising from the would-be Goldstone bosons. Note that the loop
integral goes to zero when $v_D$ goes to zero, as expected from
gauge invariance.

This operator allows collisions at high energy colliders to produce (multi-particle) dark matter states, and
is bounded by searches for mono-jets recoiling against missing momentum \cite{Goodman:2010ku,Bai:2010hh}.  While
detailed analyses for this specific interaction do not exist, existing mono-jet searches are expected to require $M_* \gtrsim $~a few hundred GeV
\cite{ATLAS:2012ky}.

%%%%%%%%%%%%%%%%%%%%%%%%%%%%%%%
\section{Effective Local Theta}
\label{sec:dm}
%%%%%%%%%%%%%%%%%%%%%%%%%%%%%%%

As we will see below, the necessary masses for the dark matter are very small, and we assume that the
local dark matter in the galactic halo can be described as a coherent state characterized by its expectation values of
energy and the quantity $\langle S^{0ij} S_{0ij} A^{2n} \rangle$ contained in the interaction Eq.~(\ref{eq:interaction}).  
These two quantities are simultaneously measurable, as can be demonstrated by observing that the Hamiltonian density
$\mathcal{H} \equiv T^{00}$ is the
$00$ component of the energy momentum tensor, which in the noninteracting limit takes the form
$T^{\mu\nu} =  F^{\mu \alpha}F_{\alpha}^{\nu} + \frac{1}{4}\eta^{\mu \nu}F^{\rho \sigma}F_{\rho\sigma} + m^2(A^{\mu}A^{\nu} - \frac{1}{2}\eta^{\mu\nu}A^{\rho}A_{\rho})$,
and satisfies $[S^{0 i j},\mathcal{H}] = 0$.  In the non-relativistic limit, $\mathcal{H}$ reduces to $m^2 A^2$, such that
$S^{0ij} S_{0ij} \mathcal{H}^n / m^{2n} \rightarrow S^{0ij} S_{0ij} A^{2n}$.

The dynamics of the dark matter in a region of space close to the solar location is described by a
partition function with the UV dynamics of QCD
encoded (schematically) by a short distance potential and the long distance influence of the gravitational dynamics of the galaxy
represented by an external potential:
\bea
- \Lambda^4 \cos \left( \frac{S^{\mu \nu \rho} S_{\mu \nu \rho} \left(A^2 \right)^n}{M^{(6+2n)}_*} - \overline{\theta} \right) - \mu ~T^{00} ,
\label{eq:potential}
\eea
with $\mu$ adjusted such that it enforces the local energy density consistent with the Galactic gravitational dynamics,
\bea
\langle T^{00} \rangle = \rho_\odot \sim 0.3~{\rm GeV} / {\rm cm}^3 \sim 3 \times 10^{-7}~{\rm eV}^4.
\eea

In a particular region of space, the contribution from the dark matter to the effective $\theta$-term is bounded by the maximum spin density consistent with the
local number density of the dark matter.  In terms of the amplitude of the coherent state ${\cal A}$, the derivatives scale as 
$\langle \partial_0 A \rangle \sim m {\cal A}$,
$\langle \partial_i A \rangle \sim m v {\cal A}$ (where $v \sim 10^{-3}$ is the typical velocity dispersion), 
and $\langle S^{0ij} \rangle \sim s m {\cal A}^2$, where $0 \leq s \leq 1$ characterizes
the degree to which the field is polarized.  In this language, the long distance contribution to the effective potential determines
${\cal A}$, and the QCD contribution acts to prefer a local value of $s$ which minimizes the effective $\theta$
term in that region of space.

\begin{figure}[t]
\begin{center}
\includegraphics[width=1\linewidth]{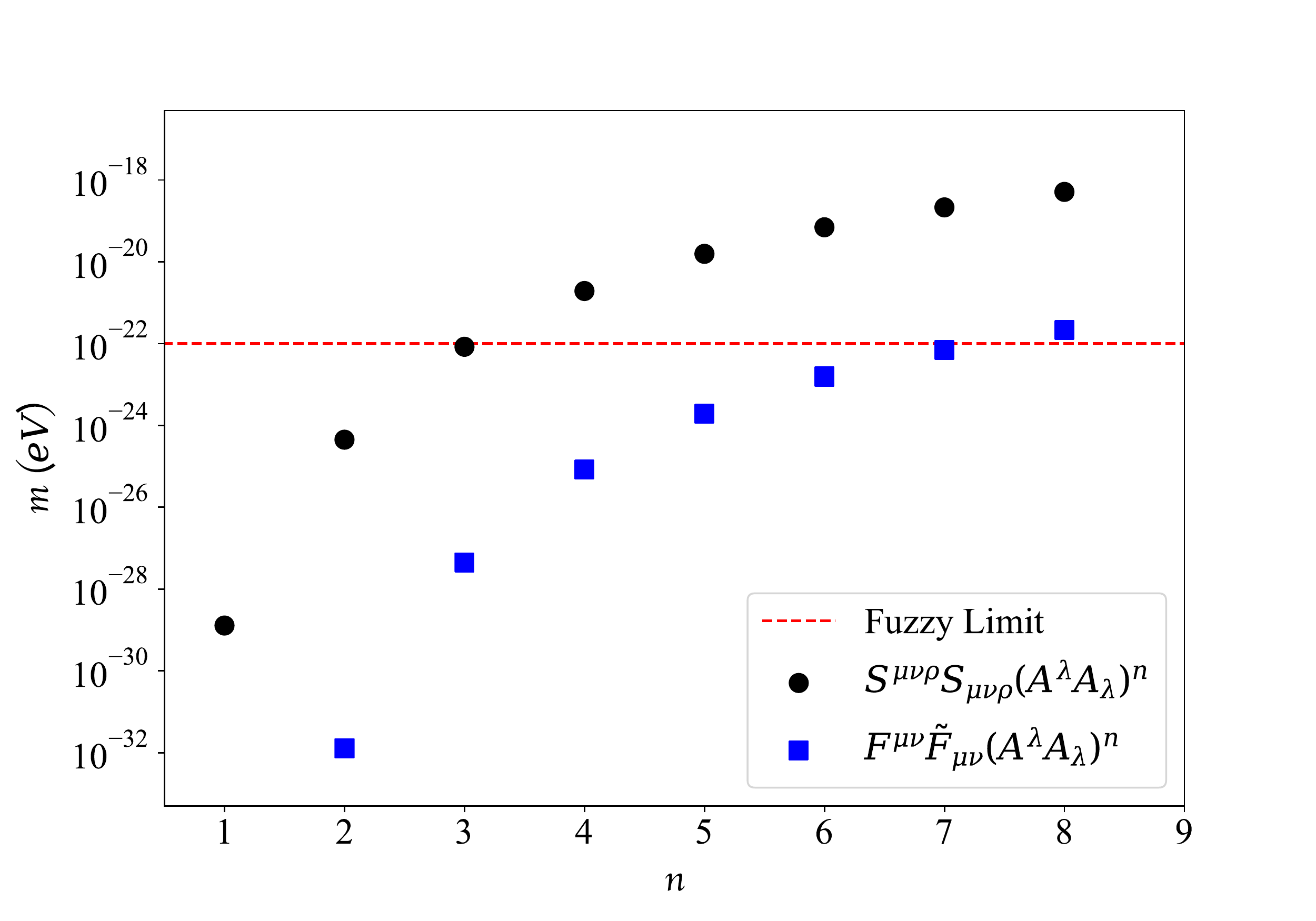}
\end{center}
\caption{Maximum dark matter mass consistent with solving the strong CP problem near the Earth, as a function of the operator dimension $n$ (black circles).  The red dashed line indicates the bound on the dark matter mass from small scale structure 
\cite{Bar-Or:2018pxz,Marsh:2018zyw,Nebrin:2018vqt}. The blue squares indicate the maximum masses from the alternative
interaction, Eq.~(\ref{eq:alternate}).}
\label{fig:mass}
\end{figure}

The dark matter contribution to the effective $\theta$ is parametrically,
\bea
\frac{ s^2 m^2 {\cal A}^{(4+2n)}}{M_*^{(6+2n)}}
\sim s^2 \frac{\rho^{(2+n)}}{M_*^{(6+2n)} m^{(2+2n)}} .
\eea
In order to cancel a $\overline{\theta}$ of order one near the Sun, the mass of the dark matter must satisfy,
\bea
m \lesssim \left( \frac{\rho_\odot^{(2+n)}}{M_*^{(6+2n)}} \right)^{\frac{1}{2+2n}}.
\eea
The maximum $m$ as a function of the operator dimension $n$ is plotted for $M_* = 1$~TeV in Figure~\ref{fig:mass}.  
For $n \geq 3$, masses
large enough to be consistent with the bound on the fuzziness of dark matter on small scales \cite{Bar-Or:2018pxz,Marsh:2018zyw,Nebrin:2018vqt,Nadler:2019zrb} 
are consistent with the emergent solution to the strong CP problem.

While operators containing larger values of $n$ are necessary to consistently cancel $\overline{\theta}$ near the Earth, 
it is clear that the (unavoidable) presence of operators with lower $n$ are not problematic.  Given the local density of dark matter,
the lower $n$ operators make a negligible contribution to the local effective $\theta$ term.  Operators with higher $n$ occur
at the same order in the loop expansion, though they are suppressed by additional powers of $M_*$. 

%%%%
\subsection{Additional Contributions to $\theta_{\rm eff}$}
%%%%

Our analysis so far has assumed that the QCD potential represents the only important dynamics influencing the dark matter
spin density.  It is crucial that any other contributions be sufficiently subdominant that they deflect $s$
from the minimum of Equation~(\ref{eq:potential}) such that the effective $\theta$ term remains $\lesssim 10^{-10}$.

The same dynamics which gives rise to the operator connecting the dark matter to $G \widetilde{G}$ will also lead to operators
containing dependence on $s$ which is unaligned with $\overline{\theta}$.  These operators take the form,
\bea
\frac{a_p}{16 \pi^2}\frac{1}{M_*^{(8+2p)}} \left(S^{\mu \nu \rho} S_{\mu \nu \rho} \right)^2 \left (A_\lambda A^\lambda \right)^{p}
\eea
where $p$ is an integer which characterizes the operator making the dominant contribution, and $a_p$ is a dimensionless coefficient which could be computed
given a more concretely realized UV theory.  This operator will shift $s$ from the minimum cancelling $\overline{\theta}$, inducing an
effective $\theta$ term of order:
\bea
\delta \theta & \sim & 
\frac{\rho_\odot^2}{\Lambda^4 m^2 M_*^2} \times \left( \frac{\rho_\odot}{m^2 M_*^2} \right)^{p-n} .
\eea
For $m \sim 10^{-18}$~eV and $M_* \sim 1$~TeV, the effective local $\theta$ term is
acceptably small provided $p \lesssim n+5$.

The local environment may also impose a preference on the net dark matter spin density.  For example, the dark matter
may possess a magnetic dipole moment, described by e.g.,
\bea
\frac{e \lambda_m}{16 \pi^2 M_*^4} F_{\rm EM}^{\mu \nu} ~ \partial^2 \left( F_{\nu \rho} \right)  ~ F^{\rho}_{\mu}
\eea
where $F_{\rm EM}$ is the electromagnetic field strength, $e$ is the electric coupling, and $\lambda_m$ is a dimensionless quantity.  If the mediator
fermions carry electroweak charge, one would expect the magnetic dipole is induced at one loop, and $\lambda_m \sim 1$, whereas
if not it will nonetheless be induced at three loops, $\lambda_m \sim (\alpha_S(M_*) / 4 \pi)^2$.  At the surface of the Earth,
this induces a shift in the effective theta term of order,
\bea
\delta \theta & \sim &
\frac{e \lambda_m}{32 \pi^2} \frac{B_\oplus m^{(3+n)} M_*^{(-1+n)}}{\Lambda^4 \rho_\odot^{n/2}}
\eea
where $B_\oplus \sim 3 \times 10^{-3}$~eV$^2$ is the strength of the Earth's magnetic field at its surface.  Even for $\lambda_m \sim 1$, this is far too small
to be important for the masses of interest.

If the dark matter interacts directly with electrons with coupling $g_D$
(e.g. through a small amount of kinetic mixing with the hypercharge interaction), it will 
typically induce a magnetic moment that is larger by $\lambda_m \sim g_D^2 M_*^2 / m_e^2$, where $m_e$ is the mass of the electron.  Even for 
order one coupling strengths $g_D \sim 1$, this is small enough as to not significantly destabilize the local effective value of $\theta$.

Even in the absence of a magnetic moment, there is a gravitational interaction between the dark matter spin and the spin of the Earth.  These corrections
are encapsulated by the potential on the net dark matter spin density induced by the Earth's gravitational field, described as a background Kerr 
metric characterized by its Schwarzschild radius $r_s = 2 G M_\oplus \sim 10^5$~eV$^{-1}$ 
and angular momentum per unit mass $\vec{a} = \vec{J}_\oplus / M_\oplus$; $| \vec{a} | \sim 10^{5}$~eV$^{-1}$.  
To linear order
in $r_s$ and $\vec{a}$, the term in the effective Lagrangian at a position $\vec{r}$ from the center of the Earth reads,
\begin{equation}
    \frac{r_s m}{2 r^3} (\vec{r}\times \vec{a})\cdot \{(\vec{A}\times \vec{\partial})\times \vec{A}\} + 
    \frac{r_s m^2}{r^3} A_0 ~\vec{a}\cdot(\vec{r}\times \vec{A}).
\end{equation}
The correction to the local value of the effective $\theta$ is,
\bea
\frac{r_s | \vec{a}| v}{R^2_\oplus} \frac{M_*^{3+n} m^{1+n}}{\Lambda^4 \rho_\odot^{n/2}},
\eea
where $R_\oplus$ is the radius of the Earth.  For the parameters of interest, this is negligibly small.

%%%%%%%%%%%%%%%%%%%%%%%%%%%%%%%
\section{Phenomenology}
\label{sec:pheno}
%%%%%%%%%%%%%%%%%%%%%%%%%%%%%%%

%%%%
\subsection{Cosmological Production}
%%%%

As with any ultralight boson playing the role of dark matter, it is necessary to invoke a nonthermal production mechanism which results
in a non-relativistic momentum distribution.  For the low masses of interest here, production through inflationary fluctuations
is thought to be inefficient given the current upper bound on the inflationary scale \cite{Nelson:2011sf,Arias:2012az,Graham:2015rva}.  Production through a
generic tachyonic instability is possible, though it requires some fine-tuning \cite{Co:2018lka,Bastero-Gil:2018uel,Agrawal:2018vin}.  Masses
as low as $\sim 10^{-18}$~eV can be accommodated if the vector mass results from a dark Higgs whose mass is close to the dark matter
mass \cite{Dror:2018pdh}.

%%%%
\subsection{Structure of Galaxies}
%%%%

For masses close to the fuzzy limit, small scale structures are prevented from forming, and the cusps of large galaxies are typically
smoothed into cores \cite{Hu:2000ke,Hui:2016ltb}.  For masses on the larger end of the range we consider, these effects are
unlikely to be observable.

A potentially important feature stems from the fact that dense areas
of dark matter have a smaller effective $\theta$, and thus a lower vacuum energy.  If one treats the background of dark energy
as a cosmological constant, and tunes its value such that in regions with very little dark matter, the net vacuum energy
reproduces the observed acceleration of the cosmological expansion, this implies that regions containing over-densities of
dark matter experience a net negative contribution to their vacuum energy from QCD.  This feature
could lead to interesting modifications to the usual cosmology and history of structure formation (e.g. \cite{Grossi:2008xh}).
However, at face value this picture implies a dramatic modification to the dynamics of galaxies, and may pose a serious
challenge unless there is some mechanism which operates locally to cancel contributions to dark energy 
(perhaps as a solution to the cosmological constant problem).

A less dramatic solution would be to invoke $n \gtrsim 6$ and dark matter masses closer to the fuzzy limit, for which the cosmological
density of dark matter is sufficient to solve the strong CP problem across the entire Universe.  In that case, one adjusts the
cosmological constant such that it leads to the observed cosmological acceleration, without any particular impact on galactic dynamics.

%%%%
\subsection{Signals at Gravitational Wave Detectors}
%%%%

The mechanism by which the vector dark matter environmentally solves the strong CP problem is somewhat agnostic as to its interactions with the
Standard Model fermions.  There could be a small direct coupling, or one could be induced through kinetic mixing with the ordinary photon.  In that
case, the motion of the Earth through the dark matter halo induces an additional time-dependent contribution to the force between objects
at a tiny level which is nonetheless accessible to interferometers designed to detect gravitational waves \cite{Pierce:2018xmy}.
In the mass range of interest, the current best constraints from the E\"{o}t-Wash experiment \cite{Su:1994gu,Schlamminger:2007ht}
require the coupling to ordinary matter be less than about $e \times 10^{-23}$, depending on the details of which SM fermions interact with
the light boson, and the LISA experiment is expected to eventually improve on these limits for masses
$\gtrsim 10^{-18}$~eV \cite{Pierce:2018xmy}.

%%%%
\subsection{Distant CP Violation}
%%%%

Any environmental solution to the strong CP problem based on the background of dark matter can have an important consequence:
regions without dark matter may be unable to completely cancel the effective $\theta$, and thus have different microscopic physics
compared with the solar system, characterized by the protons and neutrons in those regions of space possessing large
electric dipole moments whose magnitude corresponds to the local value of $\theta_{\rm eff}$ 
and can be estimated from chiral perturbation theory \cite{Srednicki:2007qs,Mereghetti:2010kp},
\bea
d_p \simeq \frac{e g_A c_+ \tilde{m}~\theta_{\rm eff}}{8 \pi^2 f_\pi^2} \log \left( \frac{\Lambda^2}{m_\pi^2} \right),
\eea
where the axial coupling $g_A \sim 1.27$ and $c_+ \sim 1.7$ are terms in the chiral Lagrangian, and
 $\tilde{m} \equiv m_u m_d / (m_u + m_d) \sim 1.2$~MeV is the reduced quark mass.
In regions with $\theta_{\rm eff}$ of order one, $d_p$ is of order $10^{-16} e$~cm.
This large CP violation is unlikely to lead to large changes in stellar dynamics and 
evolution \cite{Ubaldi:2008nf},
but could potentially lead to observable deviations in the atomic physics of stars in regions with lower dark matter density, such as in
the outskirts of the Milky Way, or in nearby globular clusters.

Since the bulk composition of stars is hydrogen, we examine the impact of a proton electric dipole moment on its atomic transitions.
Treating the electric dipole as a perturbation, the first order correction to the $nlm$ electronic wave function of a hydrogen atom,
$| \delta\Psi_{nlm} \rangle$, is given by,
\begin{equation}
    |\delta\Psi_{nlm}\rangle = \sum_{(n'l'm')} \frac{\langle\Psi_{n'l'm'}| \hat{H}' |\Psi_{nlm}\rangle}{E_{nlm}-E_{n'l'm'}}|\Psi_{n'l'm'}\rangle
\end{equation}
where $\hat{H}'$ is the additional electric dipole field induced by the proton at the origin, and $E_{nlm}$
and $| \Psi_{nlm}\rangle$  are the unperturbed
energy level and unperturbed state vector of the $nlm$ state.

The dipole interaction induces mixing between the unperturbed
$l=0$ and $l=1$ states, which allows for E1 single photon $2s \rightarrow 1s$ transitions through the correction to 
$|\delta\Psi_{200}\rangle$ proportional to $|\Psi_{n'10}\rangle$:
\bea
    & & \langle \Psi_{n'10} | \delta\Psi_{200}\rangle 
    =  \frac{d_p ~e}{4\pi \sqrt{3} \epsilon_0} \frac{C_{n'1}C_{20}}{E_{n'10}-E_{200}}  \\
    & & \hspace*{1cm}
    \times \int_0^{\infty} dr ~e^{-\frac{r}{a_0}(\frac{1}{n'}+\frac{1}{2})}\frac{2 r }{n' a_0}L_{n'-2}^{3} \left(\frac{2r}{n' a_0}\right)
   L_1^1\left(\frac{r}{a_0} \right)
   \nonumber
\eea
where $C_{nl}$ are the hydrogen wave function normalization coefficients, 
$a_0$ is the Bohr radius,
$L_n^l(x)$ are the associated Laguerre polynomials,
and the $z$ axis has been chosen along the direction of the electric dipole.
%and $Y_l^m(\theta, \phi)$ are the spherical harmonics.
%The dominant $l=1$ state mixing with $|\Psi_{200}\rangle$ through the electric dipole perturbation is $|\Psi_{210}\rangle$, owing to its quasi-degenerate energy eigenvalue.

The rate for E1 emission of a single photon via the transition from the $2s$ to the $1s$ state is \cite{Sakurai:1967},
\bea
\Gamma (2s \rightarrow 1s + \gamma ) &=&
    \frac{e^2 \omega^3}{3 \pi} | \langle \Psi_{100}| ~ \hat{r}~ |\delta\Psi_{200}\rangle |^2 \\
     & \simeq & 10^{-24}~{\rm eV}\times \theta^2_{\rm eff} ,
\eea
where $\hat{r}$ is the position operator and $\omega \equiv E_{200} - E_{100}$.
In regions where $\theta_{\rm eff}$ is of order unity, this represents an enhancement of the rate
for this transition by a factor of about $10^{4}$ compared with the CP-conserving
M1 transition \cite{Hitoshi}.  In principle, a
powerful telescope collecting spectroscopic information could potentially discern this transition line
and infer its rate.  Resolving this transition from the nearby
CP-conserving $2p \rightarrow 1s$ line would require a wavelength resolution of order 
$\delta \lambda / \lambda \sim 10^{6}$, which is about an order of magnitude beyond the
current capabilities of an instrument such as the Keck telescope \cite{Vogt:1995zz}.

%%%%%%%%%%%%%%%%%%%%%%%%%%%%%%%
\section{Conclusions and Outlook}
\label{sec:conclusions}
%%%%%%%%%%%%%%%%%%%%%%%%%%%%%%%

We have explored a novel solution to the strong CP problem based on the dark matter environment.
The dark matter is an ultralight light vector particle with mass $\lesssim 10^{-18}$~eV, whose spin density is coupled to the
gluon field in such a way as to allow it to cancel an order one $\overline{\theta}$ at the position of the Earth.  Regions with 
sufficiently small densities of dark matter cannot locally cancel an order one $\overline{\theta}$, perhaps leading to areas of
the Universe in which CP is not locally conserved, and potentially a novel history for structure formation.

We have explored a particular operator, Eq.~(\ref{eq:interaction}), in which the dark matter spin is coupled to the gluon
$G \widetilde{G}$.  There are a wider array of possible operators, as any operator involving the dark matter spin (and
enhanced by its number density) could potentially work.  For example, the operator,
\bea
\frac{\alpha_s}{16 \pi} \frac{1}{M_*^{(4+2n)}} ~ F^{\mu \nu} \widetilde{F}_{\mu \nu} \left(A^\lambda A_\lambda \right)^n ~ 
G^a_{\sigma \lambda} \widetilde{G}_a^{\sigma \lambda}
\label{eq:alternate}
\eea
is less suppressed by the interaction scale $M_*$, though additionally suppressed from 
the spatial derivatives of the dark matter field.  From Figure~\ref{fig:mass}, we see that slightly lower masses for the
dark matter, though nonetheless consistent with the fuzzy limits for $n \gtrsim 7$, 
are required to cancel an order one $\overline{\theta}$
at the position of the Earth.  This operator has the additional complication that 
$F \widetilde{F} A^{2n}$ does not commute with the Hamiltonian, implying an intrinsically
quantum mechanical dynamic for the evolution of the Galaxy.  We leave more detailed thought concerning this
interesting possibility for future work.

\section*{Acknowledgements}

We are grateful for discussions with Aaron Barth, Matt Buckley, James Bullock, Linda Carpenter, Raymond Co, Michael Gellert,
Graham Kribs, T.C. Yuan,
and especially Chanda Prescod-Weinstein. 
This work is supported in part by NSF Grant No.~PHY-1620638, and was performed in part at Aspen Center for Physics, 
which is supported by National Science Foundation grant PHY-1607611. 

\bibliography{ref}

\end{document}